\documentclass[]{spie}  

\usepackage{amsmath,amsfonts,amssymb,siunitx}
\usepackage{graphicx}
\usepackage[colorlinks=true, allcolors=blue]{hyperref}

\title{High-statistics simulations of NewAthena WFI background using Geant4}

\author[a]{Matthew K. Heine}
\author[a]{Catherine E. Grant}
\author[a]{Marshall W. Bautz}
\author[a]{Beverly J. LaMarr}
\author[a]{Eric D. Miller}
\author[b]{Michael W. J. Hubbard}
\author[b]{David Hall}
\author[b]{Joan Requena}
\author[c]{Emanuele Perinati}
\author[d]{Steven W. Allen}
\author[d]{Artem Poliszczuk}
\author[e]{Dan Wilkins}
\author[f]{Fabio Gastaldello}
\author[f]{Silvano Molendi}
\author[g]{Ralph P. Kraft}
\author[g]{Gerrit Schellenberger}
\author[a,h]{Arnab Sarkar}
\affil[a]{Massachusetts Institute of Technology, Cambridge, Massachusetts, USA}
\affil[b]{The Open University, Milton Keynes, UK}
\affil[c]{Universit\"at T\"ubingen, Institut f\"ur Astronomie und Astrophysik, T\"ubingen, Germany}
\affil[d]{Stanford University, Stanford, California, USA}
\affil[e]{The Ohio State University, Columbus, Ohio, USA}
\affil[f]{INAF/IASF-Milano, Milano, Italy}
\affil[g]{Center for Astrophysics $\vert$ Harvard \& Smithsonian, Cambridge, Massachusetts, USA}
\affil[h]{University of Arkansas,
Fayetteville, Arkansas, USA}

\authorinfo{Further author information: (Send correspondence to M.K.H.)\\M.K.H.: E-mail: mheine@mit.edu}

\begin{document} 
\maketitle

\begin{abstract}
The observation of hot gas structures is one science goal of the Wide Field Imager (WFI) on ESA’s NewAthena X-ray observatory. Because the measurement of these faint diffuse sources is limited by background from cosmic ray particle interactions within the instrument, understanding and reducing this background is critical. To this end, we employ a two-pronged approach, performing high-fidelity Geant4 simulations on both detailed, realistic geometry models as well as complementary simple geometry models. The former can reveal subtle sensitivities of background to details of the instrument design. The latter allows for fast iteration, useful in guiding and understanding the larger simulations. We show how we leverage High Performance Computing (HPC) resources to achieve simultaneously high throughput and fast time to result. We discuss our recent results, which are applicable not only to WFI, but also other X-ray missions.
\end{abstract}


\section{INTRODUCTION}  
\label{sec:intro}  

ESA's next large X-ray observatory, NewAthena, is scheduled to launch to Earth-Sun L1 in the late 2030s.\cite{cruise2025newathena} NewAthena will explore the energetic and hot universe using two selectable focal plane instruments. The Wide Field Imager (WFI) will provide moderate spectral resolution over a large 40~arcminute field of view\cite{WFIref} and the X-ray Integral Field Unit (X-IFU) will provide high-resolution spectroscopy\cite{xifu}. The WFI makes use of depleted p-channel field-effect transistor (DEPFET) active pixel sensor arrays with a pixel size of 130~$\mu$m $\times$ 130~$\mu$m, which are fully depleted to 450~$\mu$m. The full frame readout time is 5~msec, and the energy range is 0.2--15~keV.\cite{WFIDEPFETref}

The science goals of NewAthena include the observation of faint diffuse sources.\cite{cruise2025newathena}  Consequently there has been significant effort for some time towards both understanding and reducing background caused by the interaction of particles and X-rays with the instrument.  These efforts include Monte Carlo simulations of instrument background to both inform the physical design of the instrument\cite{kienlin2018evaluation}, and optimize readout and telemetry strategies,\cite{grant2020reducing,ericjatis} and novel event characterization and filtering techniques utilizing machine learning\cite{wilkins_ml,artem_ml}, as well as better characterizing the spectral and temporal properties of various particle background components.\cite{Gastadello,arnab,gerrit}

In this manuscript, we will describe how we are able to leverage high performance computing (HPC) resources to perform high-fidelity, high-statistics simulations to enable the aforementioned goals of understanding and reducing the instrumental background.  Such efforts are still ongoing, but we have selected a few examples which demonstrate the capability and utility of these types of simulation-enabled investigations as well as the need for computationally expensive simulations, and thus, the necessity of effective and efficient use of HPC resources.

In section \ref{sec:g4hpc}, we describe the Geant4 simulation, our HPC resources, and how our Geant4 simulations and post-processing are deployed on this HPC system. In section \ref{sec:results} we examine example studies which demonstrate how these HPC-enabled simulations of both a detailed and simplified mass model can be used to assist design optimization and understand instrument background.

\section{GEANT4 SIMULATIONS AND HPC ENVIRONMENT}  
\label{sec:g4hpc}  

We employed a Monte Carlo simulation of particle interactions within the instrument using the Geant4 simulation toolkit.\cite{AGOSTINELLI2003250, allison2006geant4, ALLISON2016186}  More details of the Geant4 WFI instrument background simulation effort can be found in Ref.~\citenum{eraerdsjatis}.  At a high level, initial \textit{primary} particles, aka \textit{primaries}, are drawn from appropriate probability distributions of particle energy for Galactic Cosmic Ray (GCR) Protons, GCR Alpha particles, GCR Electrons, and Cosmic X-ray Background (CXB), and for an isotropic source distribution as described in Ref.~\citenum{eraerdsjatis}. Then each primary is followed where it may interact with the instrument, represented in the simulation by a \textit{mass model}.  These interactions with the instrument may create additional, non-primary, particles.  All such particles are followed, and any interactions with the detectors are logged. Realistic post-processing is applied to translate detector interactions to charge in pixels, then finally to candidate X-ray events. From this we produce background spectra as well as trace sources of background within the instrument.  Such simulations are leveraged to investigate and optimize the impacts of instrument design choices on background both quantitatively and by uncovering qualitative insights.  They are also used to gain insight towards improving the reduction of background via post-processing, and again, to quantitatively evaluate different post-processing algorithms.

A consideration of the energy scales and relevant physics processes dictates that such a use case requires these simulations to include accurate electromagnetic modeling as well as atomic relaxation processes.  Most of our simulations are performed using the \verb|QBBC_EMZ| physics lists with fluorescence,  Auger processes, particle-induced X-ray emission (PIXE), and atomic de-excitation enabled. 
All results shown in this paper were obtained using Geant4 version 11.4.0.

\begin{figure}[!htbp]
    \centering
    \includegraphics[width=0.75\linewidth]{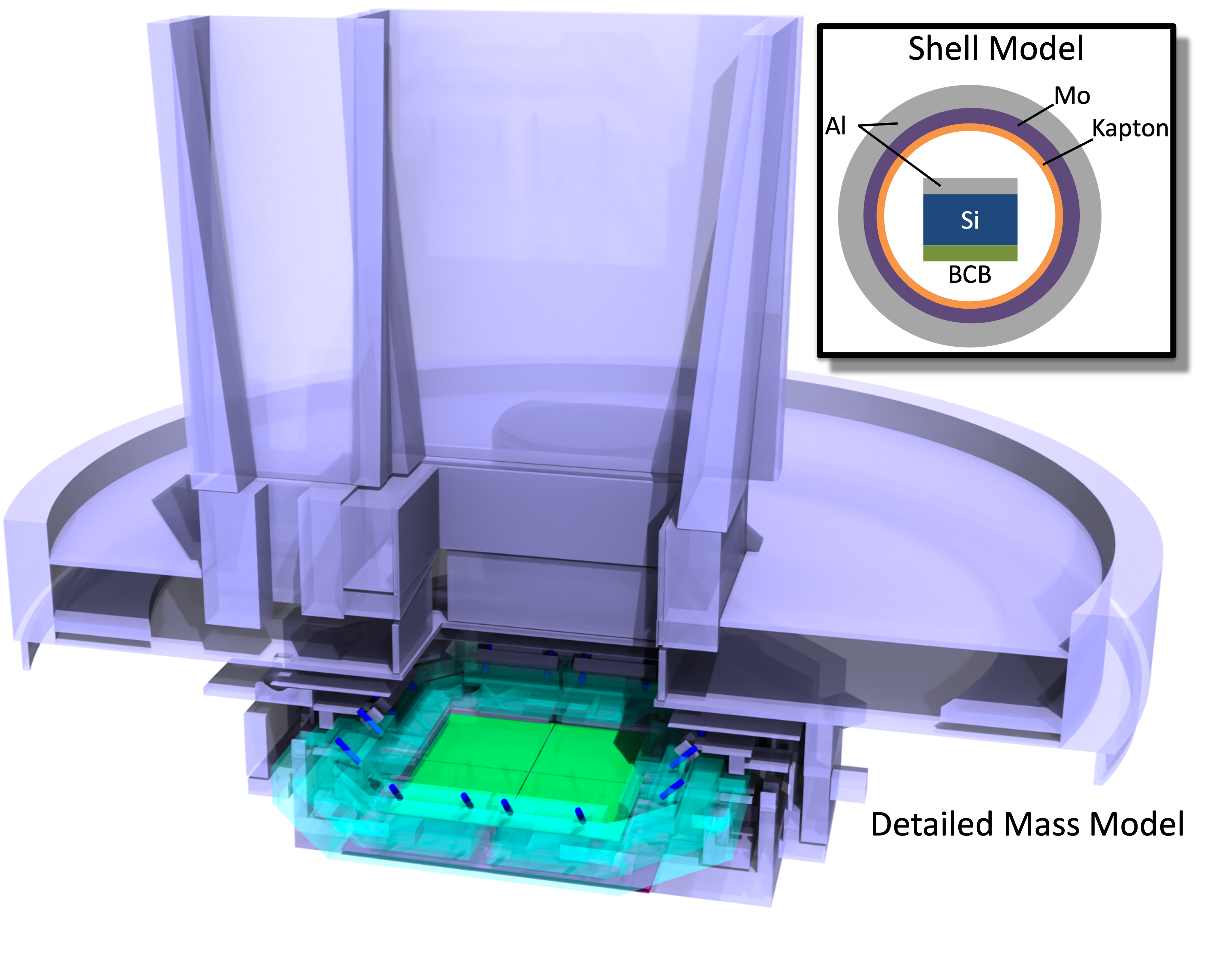}
    \caption{A cutaway view of the detailed mass model.  A few features are highlighted: the DEPFETs are shown in green, the bolts in blue, and the molybdenum structure surrounding the DEPFETs are shown in cyan.  Components are rendered semi-transparent to allow more of the mass model to be visible. (The model shown here is an intermediate working model, not the final WFI design.) Inset: Shell mass model, which is a simplified model consisting of a silicon slab with aluminum and BCB coatings inside aluminum, molybdenum, and Kapton concentric spherical shells.}
    \label{fig:massmodel}
\end{figure}

As implied above, the term mass model refers to the specific geometry including materials used to model the instrument in the Geant4 simulations.  The mass model is customizable and defined in a GDML format.  We employ two main flavors of mass models: the ``detailed mass model" and the simple ``shell model", both depicted in Fig.~\ref{fig:massmodel} as well as compared and contrasted in Table \ref{tab:shell}.

\begin{table}[ht]
\caption{Comparison of simple ``shell model" vs the ``detailed mass model".} 
\label{tab:shell}
\begin{center}       
\begin{tabular}{|p{0.25\textwidth}|p{0.3\textwidth}|p{0.3\textwidth}|}
\hline
\rule[-1ex]{0pt}{3.5ex}  \ & Shell Model & Detailed Mass Model \\
\hline
\rule[-1ex]{0pt}{3.5ex}  Use & Quickly iterate over many scenarios and extract general trends  & Accurate, detailed simulations  \\
\hline
\rule[-1ex]{0pt}{3.5ex}  Computational Expense & Faster/Cheaper simulation & Expensive simulation  \\
\hline
\rule[-1ex]{0pt}{3.5ex}  \# physical components & 6 & $\sim$ 1500 \\
\hline
\rule[-1ex]{0pt}{3.5ex} \% of primary particles producing detector interactions & $\sim$ 15\% & $\sim$ 2\% \\
\hline
\rule[-1ex]{0pt}{3.5ex} \# primary particles per typical study & $\sim$ \num{1e6} -- \num{5e7} & $\sim$ \num{1e9} \\
\hline
\end{tabular}
\end{center}
\end{table} 

The detailed mass model is derived from the instrument CAD and represents a careful balance between the inclusion of details of actual proposed designs but not excessive detail which would unnecessarily add significant computational cost.  For instance, different detailed mass models are considered, which might represent different proposed designs, and not all components are represented in all detailed mass models.  Furthermore, when detailed components are included, a slightly simplified geometry of each component may be used.  For example, below we will examine the effects of fasteners on background.  In these studies, the fasteners are represented in the mass model as simple solid cylinders.  The inclusion of threads, for instance, would add unnecessary complexity to the geometry, adding to the computational expense of the simulation without the expectation of any meaningful impact on the actual background.  Thus an investigation of a detailed mass model is achieved in that minor components such as fasteners are included, albeit in a sensibly efficient manner.  In detailed mass model simulations, the spherical surface emitting the primary particles, as described in Ref.~\citenum{eraerdsjatis}, is large enough to enclose all contents of the mass model, but translated so as to use the smallest radius which does so, for simulation efficiency.

The shell model, on the other hand, is a simplified model which consists of a single slab of Si representing the DEPFETs with a thin layer of Al coating on top (for blocking optical light) and BCB beneath (for passivation of the sensor).  This simplified detector is encased in concentric spherical shells of Al, Mo, and Kapton as an extremely simplified, spherically symmetric representation of the shielding and remainder of the instrument.  The spherical surface emitting the primary particles has a radius slightly larger than the outer radius of the Al spherical shell.  As shown in Table \ref{tab:shell}, this compact spherical geometry results in a higher yield of detector hits per primary particle resulting in higher simulation efficiency.  Coupled with reduced simulation cost for the simplified geometry, this leads to a far more efficient simulation compared to the detailed mass model, at the expense, of course, of detailed information.  Such a simplified mass model is nevertheless helpful in that it allows rapid iteration which can be helpful in extracting general trends, isolating potential sources of observed effects, or preliminary exploration of ``what if" scenarios.

Due to the high computational cost, summarized in Table \ref{tab:simsize}, we run our Geant4 \cite{AGOSTINELLI2003250, allison2006geant4, ALLISON2016186} simulations as well as post-processing in a high performance computing (HPC) environment, specifically the SuperCloud \cite{reuther2018interactive} system.  Since Geant4 does not significantly leverage GPUs, we run the simulation on CPU nodes, specifically the Intel Xeon Platinum 8260 nodes, which each contain 48 CPU cores and 192 GB RAM.  We have consistent, ``steady state" access to 8 such nodes (384 CPU cores simultaneously) and ``burst" access to 24 nodes (1,152 CPU cores simultaneously); by ``burst" access we mean a temporary increase in resource allocation made upon special request to and approval by the system administrators, intended for special high-priority and studies which require a short turnaround time (TAT).

\begin{table}[ht]
\caption{Typical simulation sizes for a detailed mass model study broken down by source primary particle type.  Quoted wall times are for high-priority ``burst" studies, ordinary priority (sustained resources) run time is 3$\times$ longer.} 
\label{tab:simsize}
\begin{center}       
\begin{tabular}{|l|l|l|l|l|} 
\hline
\rule[-1ex]{0pt}{3.5ex}  
Primary Particle &
\# Primaries & 
CPU Time &
Wall Time &
Output Size \\
\hline
\rule[-1ex]{0pt}{3.5ex}  
GCR Protons &
\num{1e9} &
4.8 CPU-months &
9 hours &
1.2 TB \\
\hline
\rule[-1ex]{0pt}{3.5ex}  
GCR Alphas &
\num{3.5e8} &
5.3 CPU-months &
10 hours &
0.9 TB \\
\hline
\rule[-1ex]{0pt}{3.5ex}  
GCR Electrons &
\num{1e9} &
4.8 CPU-months &
9 hours &
1.2 TB \\
\hline
\rule[-1ex]{0pt}{3.5ex}  
CXB X-Rays &
\num{5e10} &
2.7 CPU-months &
5 hours &
2.5 TB \\
\hline
\end{tabular}
\end{center}
\end{table} 

From a parallel workflow perspective, each primary particle Geant4 simulation is completely independent from all other primary particle simulations.  For example, a typical simulation of the effects of GCR Protons involves running \num{1e9} primary particles (protons); such a run consists of \num{1e9} completely independent Geant4 simulations.  Thus, the Geant4 simulations represent a high-throughput or perfectly-parallel workflow.  The Geant4 results are post-processed then ultimately gathered, so this could also be described as a map-reduce workflow.  While, in principle, each primary particle simulation \textit{could} be run as its own Geant4 simulation in parallel, that would not be an efficient workflow, given the duplicated overhead for each such simulation.  Rather, it is more efficient to divide the total number of simulated primaries into chunks over many CPU cores and have each core simulate a different chunk.  Thus, the primaries within a chunk are simulated serially by a given CPU core, but many CPU cores run simultaneously, so the different chunks are processed in parallel.  In our simulations, the number of primaries simulated on a single core is large enough that the simulation startup overhead for each core is negligible compared to the total simulation time.  With linear scaling, this leads to a factor of 384$\times$ speedup for normal runs and 1,152$\times$ speedup for high-priority ``burst" runs.  In other words, for high-priority runs, we can run in a single day what would take a single CPU core over 3 years to run.

Furthermore, since our environment is a HPC-style environment with 48 cores on a single node, we employ the multi-threaded version of Geant4\cite{ALLISON2016186}.  This \textit{significantly} decreases the memory footprint of the set of simulations performed on each node, as workers leverage shared memory.  Thus, for a standard steady-state (non-``burst") simulation, we run a separate multi-threaded instance of our Geant4 simulation on each node, using 48 threads per node, with random seeds carefully chosen based off the job, node, and thread identifiers as well as time, to ensure statistically independent simulations.  Through testing, we verified that increasing the thread count beyond 48, the number of CPU cores on the node, leads to significantly less than linear performance scaling, so the efficient 48 threads per node is used.  Jobs are submitted to and managed by the SLURM workload manager\cite{slurm}.  We also employ the built-in Geant4 thread-pinning feature to lock worker threads to CPU cores.

To enable rapid iterations of various scenarios as well as easy reproducibility, the key configurable parameters of the Geant4 simulations, such as mass model, physics lists, and detailed settings, are read from input files or set by macro text files.  For a typical detailed mass model simulation, all nodes are usually simulating different chunks of primary particles for the same set of input parameters.  If instead we are interested in comparing effects of different input parameters, for instance physics lists, the simpler shell mass model is often used.  In this case, fewer primary particles are simulated and different nodes are often running different sets of input parameters.  Thus, parallelization always occurs over primary particles but sometimes over input parameters as well.

Each Geant4 simulation is post-processed according to Ref.~\citenum{ericjatis}.  The HPC environment is leveraged here as well in that each aforementioned chunk of primary particles simulated by a single CPU core is also post-processed by a single CPU core.  Thus parallelization also occurs over sets of simulated primary particles, corresponding in the end to portions of exposure time.  At the end, results from all simulated primaries are gathered and aggregated together in a serial, non-parallel, step.  To reduce the load on SuperCloud's Lustre network file system \cite{reuther2018interactive}, the intermediate results of each Geant4 simulation and post-processing are written to the temporary scratch space on each node's local disk and only written to Lustre after aggregated into a single file.

\section{RESULTS AND DISCUSSION}
\label{sec:results}
\begin{figure}[!htbp]
    \centering
    \includegraphics[width=1.0\linewidth]{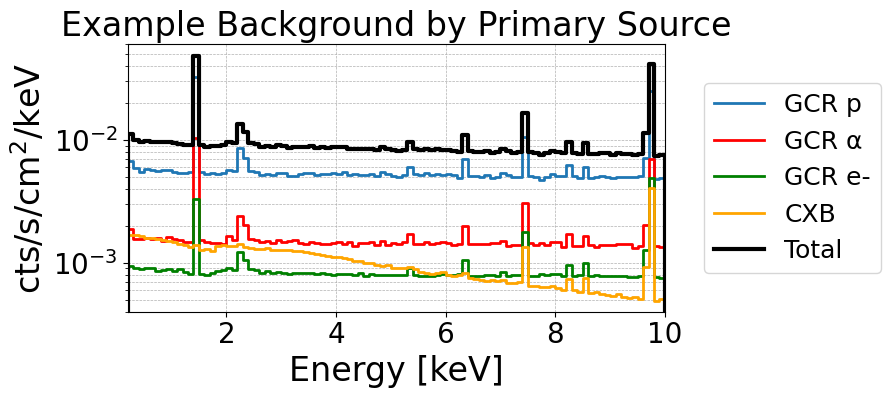}
    \caption{Example instrument background spectra produced by GCR protons, GCR alpha particles, GCR electrons, and the CXB.  The total is the sum of the four constituent spectra.  Counts are plotted on a log scale, energy is linear scale.}
    \label{fig:spectrum}
\end{figure}

An example simulated instrument background spectrum is shown in Fig.~\ref{fig:spectrum} for one particular detailed mass model.  This shows the contribution from the various primary sources: GCR protons, GCR alpha particles, GCR electrons, and the CXB, as well as a total contribution, which is modeled as simply the sum of each of the aforementioned contributions.  Here we see that GCR protons contribute most significantly to instrument background.  Overall, the shape of the various GCR-induced spectra are similar, differing mostly in overall scale.  One can also see clearly the presence of fluorescence lines in the background spectra.  These are created when particles interact with components of the instrument, resulting in the emission of characteristic X-rays for various materials.  

\begin{figure}[!htbp]
    \centering
    \includegraphics[width=0.75\linewidth]{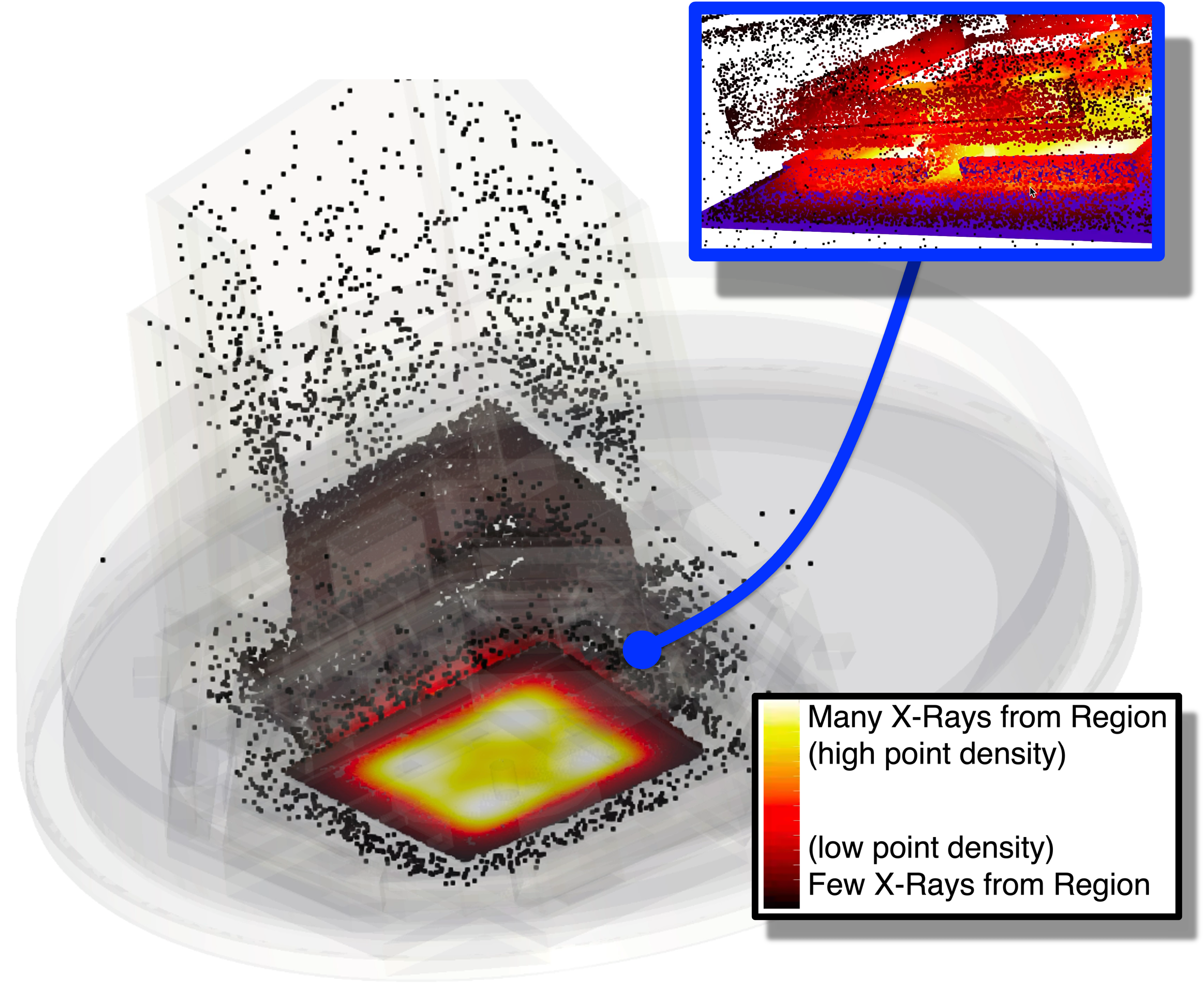}
    \caption{Origin of X-rays which enter the detector.  Mass model is shown semi-transparent in silver for reference.  Each colored point represents the initial creation point of an X-ray with energy in the 2--10~keV range which actually entered the DEPFETs.  The color of the point indicates the density of points as calculated by kernel density estimation (KDE).  Low-density points correspond to ``cooler", darker colors while high-density points correspond to ``hot" lighter colors as shown in the color bar.  High density ``hot" spots correspond to many X-rays generated in that region.  Inset shows a zoomed-in view of hot spot region near the detectors, where portions of the mass model are shown in purple for reference.}
    \label{fig:paper1}
\end{figure}

The fluorescence line component of the background spectra is, of course, highly dependent upon the details of the mass model, in particular the materials present and their geometry.  It is therefore beneficial to study and understand the sources of these fluorescence lines within the instrument, as they may be modified with changes to the instrument design.  This statement also implies that the simplified shell mass model is of limited use for such a study, so the more computationally expensive detailed mass model must be used. Additionally, since each fluorescence peak represents a very small fraction of the in-band energy window, see for example Fig.~\ref{fig:spectrum}, the simulation of many primary particles is necessary to resolve these peaks well.  We refer to simulations of many primary particles as \textit{high statistics} studies or runs.  Furthermore, the desire to clearly resolve the relative contribution of different instrument components to each fluorescence line increases the demand for higher statistics, further increasing the computational load. Consequently, HPC resources must be used to make such a study tractable. Fig.~\ref{fig:paper1} shows an example of one investigation into the spatial origins of fluorescence lines within the instrument.  For simplicity, the remainder of this study focuses only on background from GCR Proton primaries, since they represent the largest contribution to background.

\begin{figure}[!htbp]
    \centering
    \includegraphics[width=0.75\linewidth]{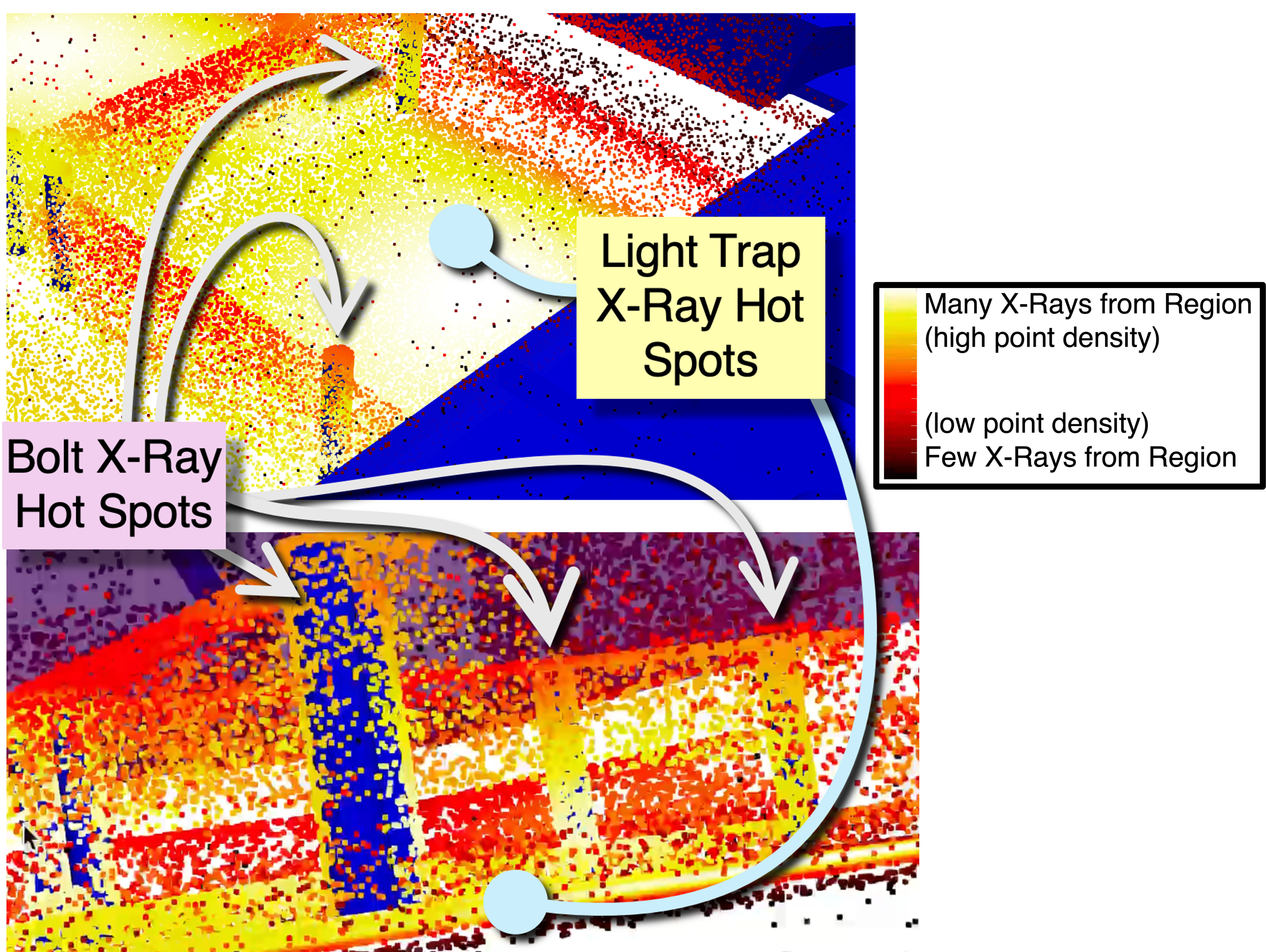}
    \caption{Origin of X-rays which enter the detector: focus on detector region.  Bolts and Light Trap physical components are shown in blue; all other components are hidden. The meaning of the colored points are described in Fig.~\ref{fig:paper1}. DEPFETs are hidden but located in the square region enclosed by the bolts and immediately above the light trap hot spots seen on the bottom.}
    \label{fig:origin}
\end{figure}

Fig.~\ref{fig:paper1} shows X-ray ``hot spots", the creation locations of in-band X-rays which enter the detector, are relatively close to the detectors themselves.  This makes physical sense given the attenuation length of X-rays in the typical materials of the mass model.  It is useful then to zoom in on the ``hot" region near the detectors, which is shown in Fig.~\ref{fig:origin}.

A close examination of Fig.~\ref{fig:origin} reveals cylindrical outlines of orange and yellow X-ray ``hot spots" in the depicted detector region.  These cylindrical regions turn out to be the exterior of the bolts within the mass model.  For a visual aid, the bolts are shown in Fig.~\ref{fig:origin} in blue.  The flat extremely ``hot" region below the bolts is a component known as the light trap.  This demonstrates not only the ability to identify components and regions of instrument background X-ray creation, but also that minor components such as bolts can contribute significantly to the X-ray background.

\begin{figure}[!htbp]
    \centering
    \includegraphics[width=0.75\linewidth]{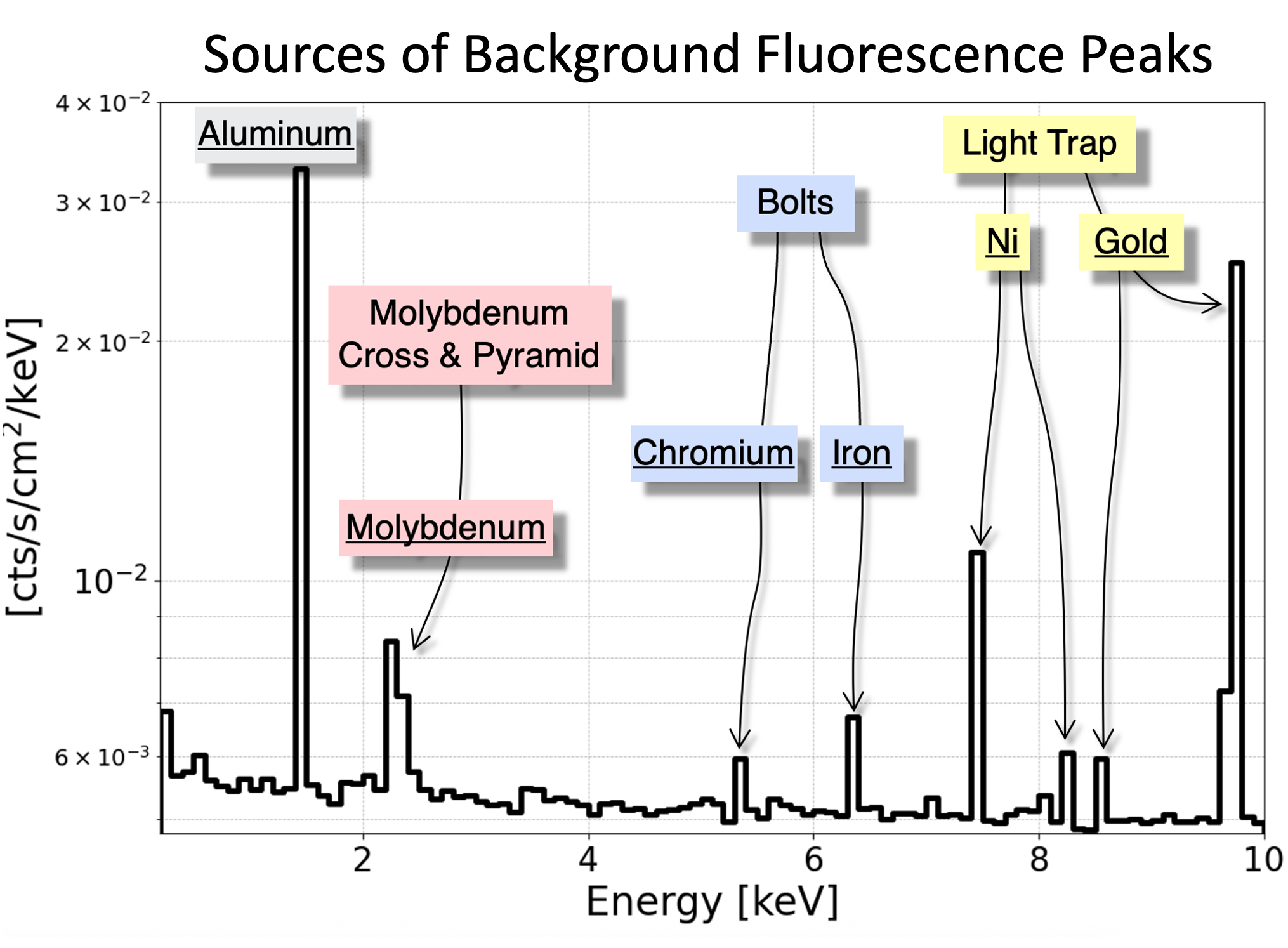}
    \caption{Sources of fluorescence peaks in simulated detector spectrum for GCR primary protons.  Simulated spectrum is for a detailed detailed mass model using Geant4 v11.4.0.  Labels correspond to the component(s) in which the X-rays of the given peak energy were predominantly found to be created in simulations.}
    \label{fig:peakid}
\end{figure}

The spatial origin information in Figs.~\ref{fig:paper1} and \ref{fig:origin} may be aggregated by tallying the points within each component to give a lower-resolution but higher-level summary of the sources of the fluorescence lines; the result is shown in Fig.~\ref{fig:peakid}, where the source components of each peak are identified.  For this mass model, it turned out that each peak, besides that of Aluminum, was caused almost exclusively by a single component, or category of components.  This need not be the case, as, in general, different components may contribute to the same peak.  Again, it is noteworthy that both the Chromium and Iron background peaks for this mass model are caused virtually exclusively by minor components, namely the bolts.

\begin{figure}[!htbp]
    \centering
    \includegraphics[width=0.5\linewidth]{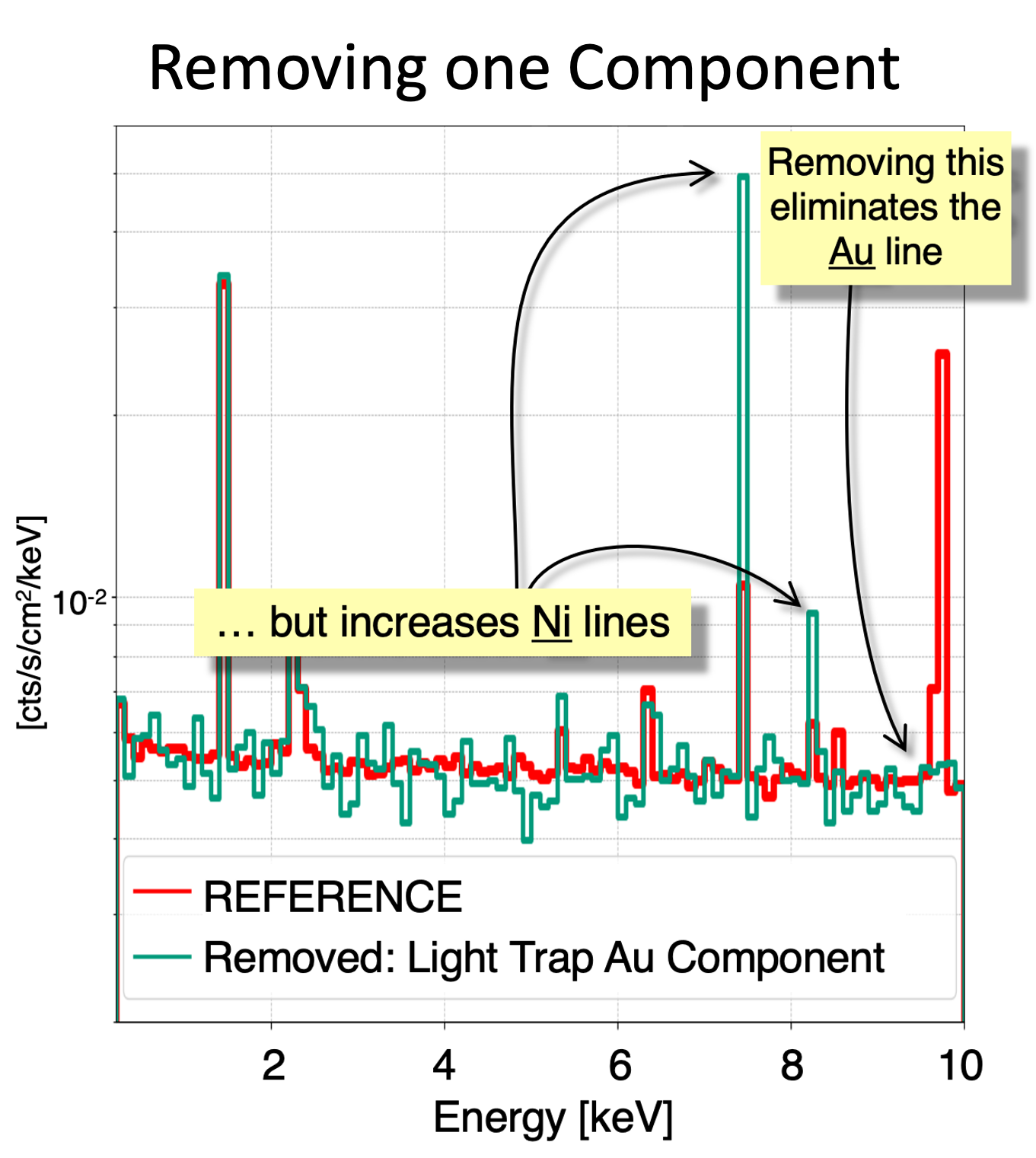}
    \caption{Example study of tradeoffs of removing a single component from the detailed mass model.  The reference spectrum, red, is that due to GCR protons shown in Fig.~\ref{fig:spectrum}. The teal spectrum is the result of a simulation where one particular gold light trap component was removed from the mass model.  The larger noise in the teal spectrum is due to the fact that far fewer primaries were simulated in the teal spectrum than the red, since the focus was merely on the effect upon the strong gold fluorescence peak.}
    \label{fig:tradeoff}
\end{figure}

Another example study, Fig.~\ref{fig:tradeoff}, is that of the tradeoffs surrounding the inclusion or removal of one particular component.  It was seen in Fig.~\ref{fig:peakid} that particle interactions with the component known as the light trap give rise to a strong gold fluorescence peak.  It turns out that this strong gold peak is caused almost exclusively by one single component within the light trap. One would expect, therefore, that removing this component would eliminate said gold peak.  Fig.~\ref{fig:peakid} demonstrates that, when this is performed in simulation, indeed the gold line is eliminated.  However, Fig.~\ref{fig:peakid} also reveals that removing this component also significantly \textit{increases} the nickel fluorescence peak.  This is because this gold light trap component actually partially shields X-rays produced within nickel elsewhere in the light trap.  This is an example of the types of tradeoff involved in design choices and the need to investigate impacts on background via high-statistics, high-fidelity simulations.

The above discussion highlights the value of detailed mass model simulations.  On the other end of the spectrum, simplified shell mass model calculations can also be of value.  These are computationally cheap, relatively speaking.  The effects of varying parameters of the mass model or the simulation itself may be quickly investigated, often with the different scenarios running simultaneously in parallel on the HPC system, yielding fast results and insights.  This is often useful for extracting general trends.  

Another useful application of the simplified shell model is investigating effects of varying Geant4 simulation settings or upgrading Geant4 versions.  With new Geant4 versions come improvements, new features, and sometimes additional physics or new abilities to tune the physics or datasets used.  The shell model provides a simplified venue in which to investigate and understand new features.  One particularly relevant example was a recent investigation into the effects of upgrading to version 11.2 of Geant4.  We noticed that, upon upgrading version 10.6.3 to version 11.2.2 of Geant4, the simulated background increased significantly.  We then employed the shell model to systematically investigate differences between these Geant4 versions in a simplified scenario, allowing us to rule out and isolate possible causes for this increased background. Ultimately, it was determined that the new General Neutron Process was the cause of this increase in simulated background. The General Neutron Process is now disabled by default in the latest version of Geant4, version 11.4, in response to bug reports.\cite{geant4_release_notes_v11_4_0}  As part of this investigation, we used the shell model to carefully investigate each new release of Geant4, including minor releases and patches, from v11.2.2 through v11.4.0. It was discussed in Ref.~\citenum{geant4training} that the General Neutron Process was intended to yield performance benefits so is not expected to quantitatively affect results to this degree, and furthermore, there is no significant additional physics captured by the process, so disabling it is expected to be safe and most reliable, without missing any new physics.  Since it was found to have a significant erroneous effect on our results, it is our recommendation to take care to disable the General Neutron Process in space applications when using a version of Geant4 prior to 11.4.

\section{SUMMARY}
\label{sec:summary}

In this proceedings manuscript we have outlined the particle background simulation pipeline we have developed for use with NewAthena WFI and its deployment in a High Performance Computing environment. This pipeline combines the Geant4 simulation toolkit for modeling the interaction of particles with the WFI instrument with realistic post-processing to mimic that done in existing X-ray astrophysics instruments. We employ a two-pronged approach, performing high-fidelity, high statistics studies when required to better understand spatial and spectral consequences of instrument design details, and lower statistics studies with simplified shell models for fast results and extracting general trends.
These results are applicable to future X-ray missions and by substituting the appropriate mass model and input particle spectra, the described pipeline can be applied to future instrumentation.

\acknowledgments
\label{sec:acknowledgments}
 
This work was done as part of the NewAthena WFI Background Working Group, a consortium including MPE, INAF/IASF-Milano, IAAT, Open University, MIT, SAO, and Stanford. We gratefully acknowledge support from NASA cooperative agreement 80NSSC21M0046 and by the the NASA Astrophysics Research and Analysis (APRA) program under grant number 80NSSC22K0342. The authors acknowledge the MIT SuperCloud and Lincoln Laboratory Supercomputing Center for providing HPC resources that have contributed to the research results reported within this paper/report.\cite{reuther2018interactive}


\bibliographystyle{spiebib} 

\end{document}